\documentclass{aa}
\usepackage{amsmath,graphics,amssymb}
\usepackage{txfonts}
\usepackage{natbib}
\newcommand{\mdot}{\ensuremath{\dot M}}
\newcommand{\Teff}{\mbox{$T_\mathrm{eff}$}}
\newcommand{\zav}[1]{\left(#1\right)}
\newcommand{\ms}{\ensuremath{\text{M}_{\odot}}}
\newcommand{\kms}{\ensuremath{\mathrm{km}\,\mathrm{s}^{-1}}}
\newcommand{\msr}{\ensuremath{\ms\,\text{yr}^{-1}}} 
\newcommand{\vel}{{\varv}}
\newcommand{\vinfty}{\ensuremath{\vel_\infty}}
\newcommand{\hzav}[1]{\left[#1\right]}

\allowdisplaybreaks

\begin{document}

\title{Hot star wind models with new solar abundances}

\author{J.  Krti\v{c}ka\inst{1} \and J. Kub\'at\inst{2}}
\authorrunning{J. Krti\v{c}ka and J. Kub\'at}

\institute{\'Ustav teoretick\'e fyziky a astrofyziky P\v{r}F MU,
            CZ-611 37 Brno, Czech Republic, \email{krticka@physics.muni.cz}
           \and
           Astronomick\'y \'ustav, Akademie v\v{e}d \v{C}esk\'e
           republiky, CZ-251 65 Ond\v{r}ejov, Czech Republic}

\date{Received 20th November 2006 / Accepted 11th January 2007}

\abstract{We compare the hot-star wind models calculated by assuming older
solar-abundance determination with models calculated using the recently
published values derived from 3D hydrodynamical model atmospheres. We show that
the use of new abundances with lower metallicity improves the agreement between
wind observation and theory in several aspects. (1) The predicted wind mass-loss
rates are lower by a factor of 0.76. This leads to better agreement with
mass-loss rates derived from observational analysis that takes the clumping into
account. (2) As a result of the lowering of mass-loss rates, there is better
agreement between the predicted modified wind momentum-luminosity relationship
and that derived from observational analysis that takes the clumping into
account. (3) Both the lower mass fraction of heavier elements and lower
mass-loss rates lead to a decrease in opacity in the X-ray region. This
influences the prediction of the X-ray line profile shapes. (4) There is better
agreement between predicted \ion{P}{v} ionization fractions and those derived
from observations.

}
\maketitle

\section{Introduction}

Massive hot stars lose a substantial part of their mass via their stellar winds.
Wind parameters, especially the wind mass-loss rate (the amount of mass lost by
the star per time unit), are the functions of basic stellar parameters, for
example the stellar luminosity \citep[see, e.g.,][for a review]{kupul, kkpreh}.
The wind mass-loss rate is the most important wind parameter, because the
presence of strong hot-star winds influences not only the emergent spectra, but
also, for example, stellar evolution \citep[e.g.,][]{vetrani}.

Since hot-star winds are mainly accelerated by the absorption (or scattering) of
radiation in lines of heavier elements (like carbon, nitrogen, or iron), the
radiative force and wind mass-loss rate also depend significantly on
metallicity. With increasing metallicity, the mass-loss rate and the wind
terminal velocity increase \citep{kupul,vikolamet,nlteii}. Frequently, the solar
abundance values of \citet{angre} are used for the wind-model predictions.
However, the solar metallicity was recently reduced following the calculations
by \citet{asgres} based on improved (self-consistent) treatment of the outer
solar convective region (using 3D hydrodynamical model atmospheres,
\citealt{nodra,stefi,asp}). The difference between older abundance
determinations and recent, more realistic ones is very striking, as the estimate
of the mass fraction of heavier elements decreased from $0.0194$ to $0.0122$.
Consequently, it would be interesting to test the influence of the new solar
abundance determination on the wind models. To do so, we compare wind models
calculated with both old and new abundance determinations.

\section{Basic assumptions}

\subsection{Wind models}

We used our own NLTE wind models to predict the hot-star wind structure \citep
[recent improvements in our calculations, especially the inclusion of the Auger
ionization, will be described elsewhere] {nltei}. Our models enabled us to
self-consistently solve the hydrodynamical equations in the
spherically-symmetric, stationary, radiatively-driven stellar wind. The
radiative force (in the Sobolev approximation, \citealt{cassob}) and the
radiative heating/cooling term were calculated using the solution of NLTE
equations, for which we mainly used the atomic data available from the Opacity
and Iron Projects \citep{topt,zel0}. This enabled us to change the abundance of
each element separately according to its new value, and not only via the common
parameter called the metallicity. Using our method we were able to predict wind
density and velocity structure, namely the wind mass-loss rate $\mdot$ and the
wind terminal velocity \vinfty.

\subsection{Model stars}

\begin{table}
\caption{Stellar parameters of the O stars.}
\label{obhvezpar}
\hfill
\begin{tabular}{rrccccc}
\hline
\hline
\multicolumn{1}{c}{Star} & \multicolumn{1}{c}{HD} & Sp. & ${R_{*}}$ & $M$ &
$\Teff$ & Source \\
& \multicolumn{1}{c}{number}& type & $[\text{R}_{\odot}]$ &$[\text{M}_{\odot}]$ &  $[\text{K}] $ \\
\hline
$\xi$ Per      &  $24912$ & O7.5IIIe & $14.0$ & $36$ & $35\,000$ & R04 \\
$\iota$ Ori    &  $37043$ & O9III    & $21.6$ & $41$ & $31\,400$ & M04 \\
15 Mon         &  $47839$ & O7Ve     &  $9.9$ & $32$ & $37\,500$ & M04 \\
               &  $54662$ & O7III    & $11.9$ & $38$ & $38\,600$ & M04 \\
               &  $93204$ & O5V      & $11.9$ & $41$ & $40\,000$ & M05 \\
$\zeta$ Oph    & $149757$ & O9V      &  $8.9$ & $21$ & $32\,000$ & R04 \\
68 Cyg         & $203064$ & O8e      & $15.7$ & $38$ & $34\,500$ & R04 \\
19 Cep         & $209975$ & O9Ib     & $22.9$ & $47$ & $32\,000$ & R04 \\
\hline
\end{tabular}
\hfill
{\em Data sources:}
M04 -- \cite{upice},
R04 -- \cite{rep},
M05 -- 
\cite{martclump}
\end{table}

For our analysis we selected O stars with $ \Teff \lesssim40\,000\,\text{K}$
(see Table\,\ref{obhvezpar}), for which the stellar parameters were derived
using the models with line blanketing. Effective temperatures and radii were
taken from \citet*[hereafter R04]{rep}, \citet*[hereafter M04]{upice}, and
\citet*[hereafter M05] {martclump}. Stellar masses were obtained using
evolutionary tracks either by us (using \citealt{salek} tracks) or by M05.

\section{Wind models with \citeauthor*{asgres} solar abundance determination}

\begin{table}[t]
\caption{Comparison of wind parameters calculated assuming
\citet{angre} and \citet{asgres} solar abundances.}
\label{zmenazpar}
\hfill
\begin{tabular}{rcccc}
\hline
\hline
\multicolumn{1}{c}{HD} & \multicolumn{2}{c}{\protect\citet{angre}}
& \multicolumn{2}{c}{\protect\citet{asgres}}\\
\multicolumn{1}{c}{number}& $\mdot$ &
\vinfty
& $\mdot$ &
\vinfty
\\
& $[\msr]$ & [\kms] &$[\msr]$ & [\kms] \\
\hline
 $24912$ & $5.7\cdot10^{-7}$ &$2440$ &$4.4\cdot10^{-7}$ &$2270$\\
 $37043$ & $7.3\cdot10^{-7}$ &$2440$ &$6.2\cdot10^{-7}$ &$2340$\\
 $47839$ & $3.7\cdot10^{-7}$ &$2450$ &$2.2\cdot10^{-7}$ &$3080$\\
 $54662$ & $1.1\cdot10^{-6}$ &$2350$ &$7.9\cdot10^{-7}$ &$2190$\\
 $93204$ & $1.8\cdot10^{-6}$ & $2490$ &$1.3\cdot10^{-6}$ &$2290$\\
$149757$ & $5.9\cdot10^{-8}$ &$2310$ &$4.7\cdot10^{-8}$ &$2040$\\
$203064$ & $7.4\cdot10^{-7}$ &$2270$ &$5.7\cdot10^{-7}$ &$2080$\\
$209975$ & $1.0\cdot10^{-6}$ &$2510$ &$8.4\cdot10^{-7}$ &$2430$\\
\hline
\end{tabular}
\hfill
\end{table}

\begin{figure}[t]
{\hfill\resizebox{0.9\hsize}{!}{\includegraphics{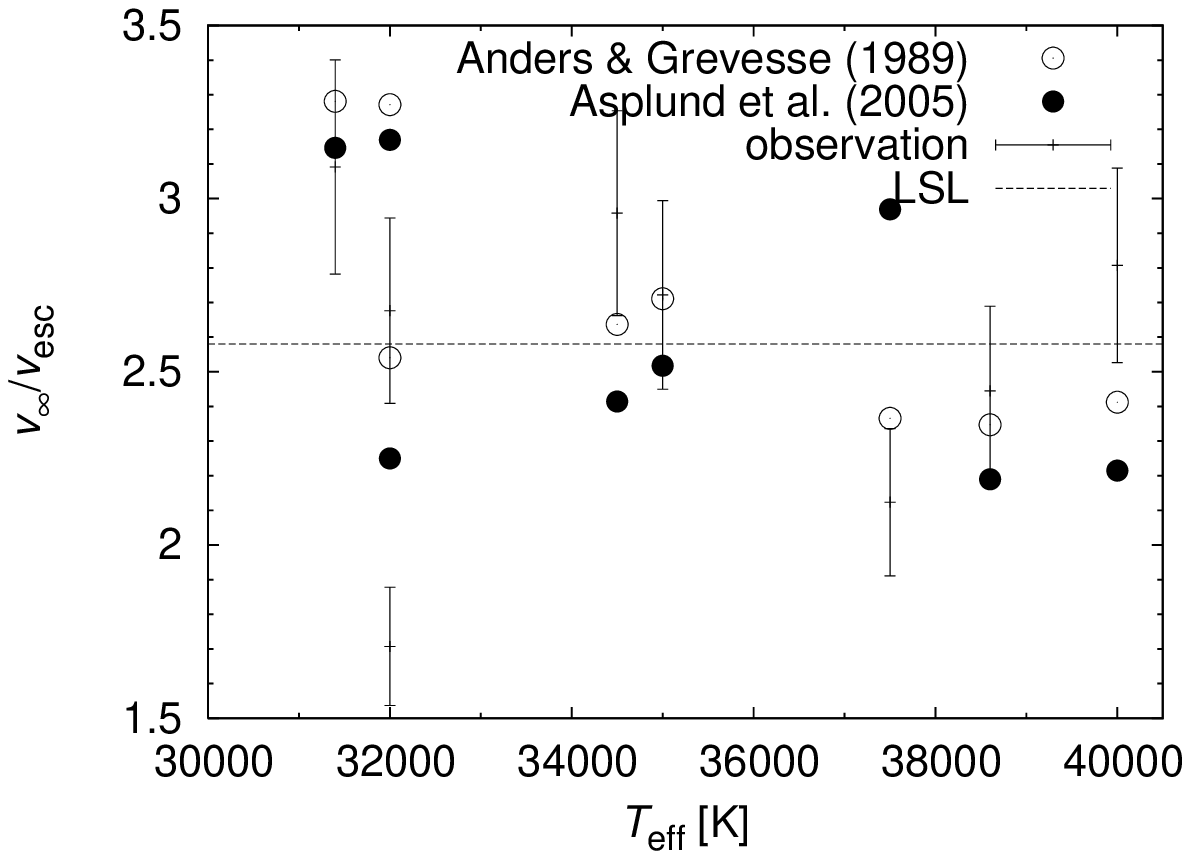}}\hfill}
\caption{Comparison of the ratio of the terminal velocity to the escape
velocity calculated using terminal velocities either from theoretical
calculations (with different sets
of abundance determinations) or derived from observations
\citep[values adopted from][M04, M05]{pulmoc}.
LSL denotes the average ratio derived by
\citet{lsl}.
}
\label{vnekvuni}
\end{figure}

With decreasing metallicity the radiative force decreases and, consequently, the
wind mass-loss rate also decreases. Moreover, the decrease in metallicity leads
to a slight lowering of the wind terminal velocity
\citep{kupul,vikolamet,nlteii}. Both these effects are apparent in
Table~\ref{zmenazpar}, where we compare wind parameters calculated using
different determinations of solar abundances (see also Fig.~\ref{vnekvuni}). On
average, the mass-loss rate changes by a factor $0.76$. This roughly corresponds
(the abundance changes were not uniform for all elements) to the dependence
$\mdot\sim Z^{0.60}$, where $Z$ is the mass fraction of heavier elements. This
dependence is slightly lower than the one reported by \citet{vikolamet} and
\citet{nlteii}. Note, however, that these results are not directly comparable,
since the relative change in the abundance of different elements is not uniform
in this study, whereas results by both \citet{vikolamet} and \citet{nlteii} were
obtained for a uniform relative change in the abundance. On average, the wind
terminal velocity also decreases with decreasing metallicity (see
Fig.~\ref{vnekvuni}). However, note that the change in the terminal velocity due
to the decrease in the metallicity is mostly lower than the error in the
terminal velocity determination.

\section{Implications of new abundance determinations}

The next logical step would be to compare our predicted wind parameters
(especially the mass-loss rates) with the quantities derived from observation.
However, such a comparison is complicated by the effect of clumping.

\subsection{Wind clumping}

There are two ways to derive wind mass-loss rates. Since many spectral features
are sensitive to wind density (or mass-loss rate), it is possible to derive
their values from observations using wind models where the mass-loss rate is a
free parameter. These values are sometimes called `observed', but we prefer to
call them `derived from observations'. On the other hand, the mass-loss rate can
be calculated {\em ab initio} from dedicated wind models where stellar
parameters (the effective temperature, mass, radius, and metallicity) serve as
input parameters. This value of the mass-loss rate is usually called
`predicted'. It is clear that both ways of determining the wind mass-loss rate
should be in close agreement. 
 
However, recent spectroscopic studies \citep{bourak, martclump, pulchuch}
indicate that there may be significant disagreement between the wind mass-loss
rates derived from observations and predicted ones. The problem is that the
emergent spectrum is, in many cases, not directly sensitive to the wind
mass-loss rate \mdot, but to the wind density squared. This means that, if it
happens that the wind is structured on small scales (``clumped''), then the
lines originating in the clumped wind with a relatively low mean wind density
may mimic those with a much higher mass-loss rate. If the winds are really
significantly clumped in all regions from which we observe the emergent
radiation, as seems to follow from the comparison of synthetic wind spectra with
the observed ones \citep[e.g.,][]{bourak,martclump}, then the real wind
mass-loss rates are significantly lower than the predicted ones.

From the studies of wind spectra that include clumping (M05), it follows that
wind mass-loss rates are roughly lower by a factor $\sim0.3$ than those derived
from the standard wind models of \citet{vikolamet}. Consequently, the reduction
of wind mass-loss rate by a factor 0.76 due to the reduced metallicity
represents an improvement in the agreement between wind mass-loss rates derived
from observation (that take the clumping into account) and predicted ones.

\subsection{Modified wind momentum-luminosity relationship}

\begin{figure}
{\hfill\resizebox{0.9\hsize}{!}{\includegraphics{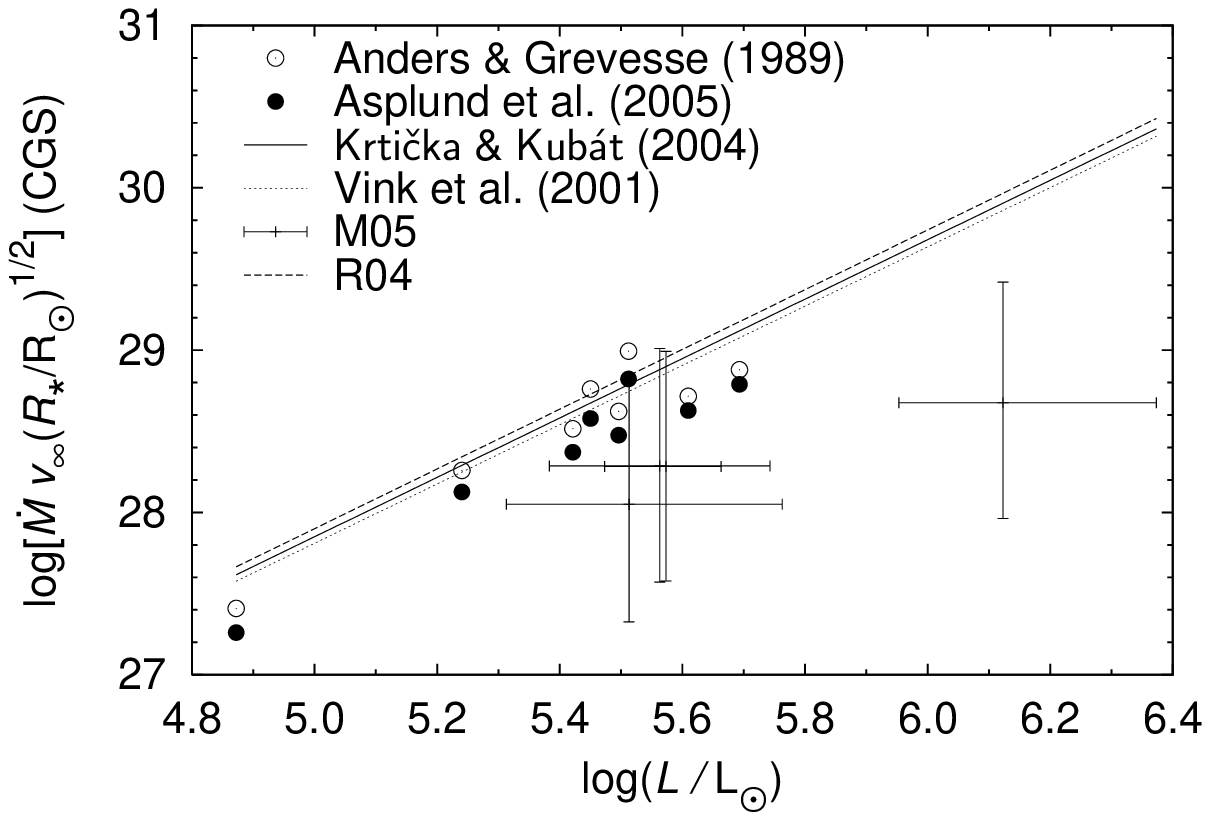}}\hfill}
\caption{Comparison of modified wind-momentum calculated for individual stars
in this paper assuming solar abundances by \citet{angre} with more recent
lower ones by \citet{asgres}. We also plot the linear fits of the theoretical 
predictions, derived with older abundances \protect\citep{nltei,vikolamet},
and values derived from observations both by M05 (individual stars
for which the clumping was taken into an account) and by R04 (linear fit for
giant and dwarfs, for which the clumping was not taken into an account).}
\label{mom_porov}
\end{figure}

It follows from both wind theory and observations that the value of the modified
wind momentum $\mdot \vinfty \zav{R/\text{R}_{\sun}}^{1/2}$ depends mainly on
the stellar luminosity \citep[and references therein]{kupul}. Especially the
mass-dependence nearly cancels out. Consequently, the wind momentum-luminosity
relationship is suitable for comparing observations and theory. Here we compare
this relationship derived using theoretical models with both older and new
abundance determinations with values from M05, who derived wind mass-loss rates
from the observed data and who account for clumping, and R04, who neglected the
possibility of wind clumping (see Fig.~\ref{mom_porov}).

The predicted wind momentum is lower for models calculated using improved solar
metallicity determination, mainly due to the decrease in wind mass-loss rate
with metallicity and, to a lesser extent, due to the lower terminal velocities
that are derived. The theoretical modified wind momentum-luminosity relationship
calculated using improved abundances has the form
\begin{equation}
\log\hzav{\mdot\vinfty\zav{R/\text{R}_{\sun}}^{1/2}}=1.91\,\log(L/\text{L}_\odot)+18.1
\quad 
\text{(CGS)}.
\end{equation}
Our results to some extent improve the agreement between the theory and
observational analysis that takes the clumping into account (M05 and also
Fig.~\ref{mom_porov}). However, they are not able to remove the difference
between the theory and observations completely. Note that there is good
agreement between theoretical predictions derived using \citet{angre} abundances
\citep[and the empty circles in the Fig.~\ref{mom_porov}] {vikolamet,nltei} and
observations by R04. Consequently, if the influence of clumping on the observed
wind spectra is negligible, then the wind models with improved abundance
determination predict lower wind momentum than those derived from observations.

\subsection{X-ray continuum optical depth}

\begin{table}[t]
\caption{Comparison of continuum opacity (opacity per unit of mass in units
of $\text{cm}^{2}\,\text{g}^{-1}$ averaged for the radius
$1.5\,\text{R}_{\odot}<r<5\,\text{R}_{\odot}$) in the X-ray region
for several different wavelengths calculated for the two 
abundance determinations (for HD~24912).}
\label{xopac}
\hfill
\begin{tabular}{lccccc}
\hline
wavelength [\AA] & 8.1 & 12.0 & 15.8 & 20.0 & 24.0\\
\citet{angre} & 31 & 83 & 141 & 228 & 154\\
\citet{asgres} & 20 & 52 & 91 & 150 & 125\\
\hline
\end{tabular}
\hfill
\end{table}

Another painful problem of the hot-star wind theory is connected with the shape
of X-ray line profiles. While the wind theory based on currently available
theoretical mass-loss rates predicts asymmetric X-ray line profiles due to the
continuum absorption within the wind \citep[see, e.g.,][]{ocpor,xlida}, the
observations mostly show symmetric X-ray line profiles. This indicates that the
wind's optical depth in continuum for X-ray frequencies is small in the region
where X-rays form.

To infer the impact of new solar abundances on the X-ray line profiles, we
compared the mass-absorption coefficients calculated with solar abundances from
both \citet{angre} and \cite{asgres}. Apparently, since heavier elements
dominate the continuum opacity in the X-ray region (especially due to Auger
transitions), the mass absorption coefficient decreases with decreasing
metallicity (see Table~\ref{xopac}). The decrease in the optical depth is not
equal for all wavelengths because different elements dominate the opacity for
different wavelengths and the relative change of the metallicity is not the same
for all elements. The influence of new solar abundance determination on the
total continuum opacity in the X-ray region is amplified by the fact that the
wind density also decreases with decreasing metallicity.

As a result of these effects, the X-ray optical depth is lower roughly by a
factor of $2$. The use of lower X-ray optical depth in the calculation of
\citet{ocpor} leads to better agreement between predicted and observed X-ray
line shapes. Conversely, the effect of scattering in these lines \citep{igrez}
might decrease with metallicity.

\subsection{Phosphorus line profiles}

The ion \ion{P}{v} is assumed to be a dominant phosphorus ion in the stellar
wind. Since its abundance is very low, its resonance lines as observed in the
hot star wind spectra are generally unsaturated. Consequently, these lines are
assumed not to be influenced by clumping and are ideal for determining the wind
mass-loss rate. This was done by \citet{fuj}, who showed that the mass-loss rate
determined from the \ion{P}{v} lines is much lower than when derived from the
H$\alpha$ line or radio emission.

This
discrepancy
can have two 
explanations: either \ion{P}{v} is not a
dominant
phosphorus
ion
or the 
winds are clumped and their
mass-loss rates are much 
lower.
However, 
the phosphorus line profiles are
sensitive to phosphorus 
abundance
and to the overall stellar metallicity (due to the dependence of wind
mass-loss rates on the metallicity).
Both these
effects
lead to the weakening of \ion{P}{v} line profiles.
The phosphorus abundance changes
by a factor of $0.81$
when using a new abundance determination
and,
together with the decrease 
in
wind mass-loss rate caused by
the decrease 
in
the
metallicity,
leads to the change 
in
\ion{P}{v} line optical depth by a factor of
about 0.62.
Although this 
shift does not
explain the main difference between theory and
observations
(possibly
also the influence of clumping
on the ionization balance
has to be accounted for,
\citealt{pulamko}), 
the use of new abundance determinations
shifts the theoretical predictions closer to
the observational results.

\section{Discussion and conclusions}

We have compared the wind models calculated using the older solar abundance
determination \citep{angre} and the solar abundances derived using 3D
hydrodynamical model atmospheres of \cite{asgres}. We have shown that the use of
new abundances (which means lowering in the metallicity, but not uniformly for
all elements) improves the agreement between observation and theory in several
different aspects:
\begin{itemize}
\item Due to lower metallicity the predicted wind mass-loss rates are lower,
consequently new models agree better with the mass-loss rates derived from
observational analysis with clumping taken into an account.
\item As a result of the lowering of mass-loss rates and slight
lowering of terminal velocities, there is better agreement between the
predicted modified wind momentum-luminosity relationship and that derived from 
observational analysis with clumping taken into account.
On the other hand, if the clumping does not significantly influence the
observed spectra, then the original good agreement between predicted
wind momentum and the one
derived from observation is worsened with models using improved solar
abundance determination.
\item Both the lower mass fraction of heavier elements and lower
mass-loss rates lead to a decrease in the opacity in the X-ray region. This 
change is the most significant one because it basically
combines the influence of two different effects
(when using predicted mass-loss rates).
\item Since both phosphorus abundance and the mass-loss rate are
lower, there is better agreement between the predicted \ion{P}{v}
ionization fraction and the one derived from observations. However, lowering 
the wind density may partly compensate for
this effect due to modification of the ionization structure.
\end{itemize}
We have to keep in mind that the actual metallicity of Galactic O stars may be
different from the solar one. Moreover, it may be modified during the stellar
evolution due to the rotational mixing \citep[e.g.,][]{mameme}, so our results
have to be understood as a manifestation of the consequences of abundance
changes in the theoretical wind models.

We can conclude that, although there is still not complete agreement
between theory and observations, the use of more realistic solar abundance
determinations may be a partial way to solve the problem because it improves
the agreement in all aspects studied here.

\begin{acknowledgements}
This research made use of NASA's ADS and the SIMBAD database, operated at the
CDS, Strasbourg, France. This work was supported by grant GA \v{C}R 205/04/1267.
The Astronomical Institute Ond\v{r}ejov is supported by the project
AV0\,Z10030501.
\end{acknowledgements}

\newcommand{\actob}{Active OB-Stars:
	Laboratories for Stellar \& Circumstellar Physics, 
 	eds. S. \v{S}tefl, S. P. Owocki, \& A.~T.
        Okazaki (San Francisco: ASP Conf. Ser),
in press}

\end{document}